\documentclass{aa}

\usepackage{graphicx}
\usepackage{txfonts}
\usepackage{color}

\begin{document}

   \title{Optical interferometry and Gaia measurement uncertainties reveal the physics of asymptotic giant branch stars}
\titlerunning{Optical Interferometry and Gaia measurement uncertainties}
%   \subtitle{I. Overviewing the $\kappa$-mechanism}

   \author{A. Chiavassa
          \inst{1}
          \and
          K. Kravchenko\inst{2}
          \and
          F. Millour\inst{1}
          \and
          G. Schaefer\inst{3}
          \and
          M. Schultheis \inst{1}
          \and
          B. Freytag \inst{4}
          \and
          O. Creevey\inst{1}
          \and
          V. Hocd\'e \inst{1}
          \and
          F. Morand \inst{1}
          \and
          R. Ligi \inst{5}
          \and
          S. Kraus\inst{6}
          \and
          J.~D. Monnier\inst{7}
          \and
          D. Mourard\inst{1}
          \and
          N. Nardetto\inst{1}
          \and 
          N. Anugu \inst{6,8}
          \and
          J.-B. Le Bouquin \inst{9,7}
          \and
          C.~L. Davies\inst{6}
          \and
          J. Ennis\inst{7}
          \and
          T. Gardner\inst{7}
          \and
          A. Labdon\inst{6}
          \and
          C. Lanthermann\inst{10}
          \and
          B.~R. Setterholm\inst{7}
          \and
          T. ten Brummelaar\inst{3}
          }

   \institute{Universit\'e C\^ote d'Azur, Observatoire de la C\^ote d'Azur, CNRS, Lagrange, CS 34229, Nice,  France \\
                \email{andrea.chiavassa@oca.eu}
         \and
         European Southern Observatory, Alonso de Cordova 3107, Santiago, Chile
         \and
         The CHARA Array, Mount Wilson Observatory, Mount Wilson, CA 91023 USA
         \and
         Department of Physics and Astronomy at Uppsala University, Regementsvägen 1, Box 516, 75120 Uppsala, Sweden
         \and
         INAF – Osservatorio Astronomico di Brera, Via E. Bianchi 46, I-23807 Merate, Italy
         \and
         University of Exeter, School of Physics and Astronomy, Stocker Road, Exeter, EX4 4QL
         \and
         Department of Astronomy, University of Michigan, Ann Arbor, MI 48109 USA
         \and
         Steward Observatory, 933 N. Cherry Avenue, University of Arizona, Tucson, AZ 85721, USA
         \and
         Institut de Planétologie et d'Astrophysique de Grenoble, CNRS, Univ. Grenoble Alpes, France
         \and
         Instituut voor Sterrenkunde, KU Leuven, Celestijnenlaan 200D, 3001 Leuven, Belgium
             }

   \date{...}

% \abstract{}{}{}{}{} 
% 5 {} token are mandatory
 
  \abstract
   {Asymptotic giant branch (AGB) stars are cool luminous evolved stars that are well observable across the Galaxy and populating Gaia data. They have complex stellar surface dynamics, which amplifies the uncertainties on stellar parameters and distances.}
   {On the AGB star CL~Lac, it has been shown that the convection-related variability accounts for a substantial part of the Gaia DR2 parallax error. We observed this star with the MIRC-X beam combiner installed at the CHARA interferometer to detect the presence of stellar surface inhomogeneities.}
   {We performed the reconstruction of aperture synthesis images from the interferometric observations at different wavelengths. Then, we used 3D radiative hydrodynamics (RHD) simulations of stellar convection with CO5BOLD and the post-processing radiative transfer code {{\sc Optim3D}} to compute intensity maps in the spectral channels of MIRC-X observations. Then, we determined the stellar radius using the average 3D intensity profile and, finally, compared the 3D synthetic maps to the reconstructed ones focusing on matching the intensity contrast, the morphology of stellar surface structures, and the photocentre position at two different spectral channels, 1.52 and 1.70 $\mu$m, simultaneously.}
   {We measured the apparent diameter of CL~Lac at two wavelengths (3.299$\pm$0.005 mas and 3.053$\pm$0.006 mas at 1.52 and 1.70 $\mu$m, respectively) and recovered the radius ($R=307\pm41$ and $R=284\pm38$ $R_\sun$) using a Gaia parallax. In addition to this, the reconstructed images are characterised by the presence of a brighter area that largely affects the position of the photocentre. The comparison with 3D simulation shows good agreement with the observations both in terms of contrast and surface structure morphology, meaning that our model is adequate for explaining the observed inhomogenities.}
   {This work confirms the presence of convection-related surface structures
    on an AGB star of Gaia DR2. Our result will help us to take a step forward in exploiting Gaia measurement uncertainties to extract the fundamental properties of AGB stars using appropriate RHD simulations.}
   
    \maketitle
%
%-------------------------------------------------------------------

\section{Introduction}

\begin{table*} 
\begin{center}
 \caption{Observational stellar properties and parameters of CL~Lac}
 \label{CLLac_tab}
 \begin{tabular}{cccccccccc}
\hline
   G\tablefootmark{a}  &  Bp\tablefootmark{a}     &   Rp\tablefootmark{a}  &  K2MASS\tablefootmark{b} & $L_\star$\tablefootmark{c} & $T_\mathrm{eff}$\tablefootmark{d}  & $p$ & $\sigma_\varpi$\tablefootmark{a} & $\sigma_\varpi$ & $P_\mathrm{puls}$\tablefootmark{g} \\
   &   &   &  &  $L_\sun$ & K & mas & mas & AU &  years \\
\hline
8.73 & 12.59 & 7.26 & 2.15 & 4888.0$\pm$8.7 & 3293 & 1.155 & 0.150 & 0.130  & 1.134 \\
\hline
 \end{tabular}
\end{center}
\tablefoot{The first four columns display the G, Bp, Rp, and K photometric colours from Gaia and 2MASS; the two following columns indicate the luminosity ($L_\star$) and effective temperature ($T_\mathrm{eff}$); the next three show the parallax ($p$) with its error ($\sigma_\varpi$) in mas and in AU; the last column reports the pulsation period ($P_\mathrm{puls}$).\\
\tablefoottext{a}{Gaia DR2 \citep{2018A&A...616A...1G}}\\
\tablefoottext{b}{2MASS \citep{2006AJ....131.1163S}}\\
\tablefoottext{c}{Computed from K2MASS with a bolometric correction of 3.0}\\
\tablefoottext{d}{The temperature is based on Gaia Bp/Rp photometric colours and was estimated for sources brighter than G=17 mag for 3000K<$T_\mathrm{eff}$<10000K. Due to the degeneracy of the photometric temperature with interstellar extinction, an empirically trained machine learning algorithm was used for external spectroscopic datasets in low-extinction regions in order to derive temperatures \citep{2018A&A...616A...8A}. However, due to the limited training set for cool stars, photometric temperatures have to be taken with extreme caution for these stars}\\
\tablefoottext{g}{\cite{1993ApJ...413..298J}}}
\end{table*}

 \begin{figure}
   \centering
    \begin{tabular}{c}
        \includegraphics[width=0.95\hsize]{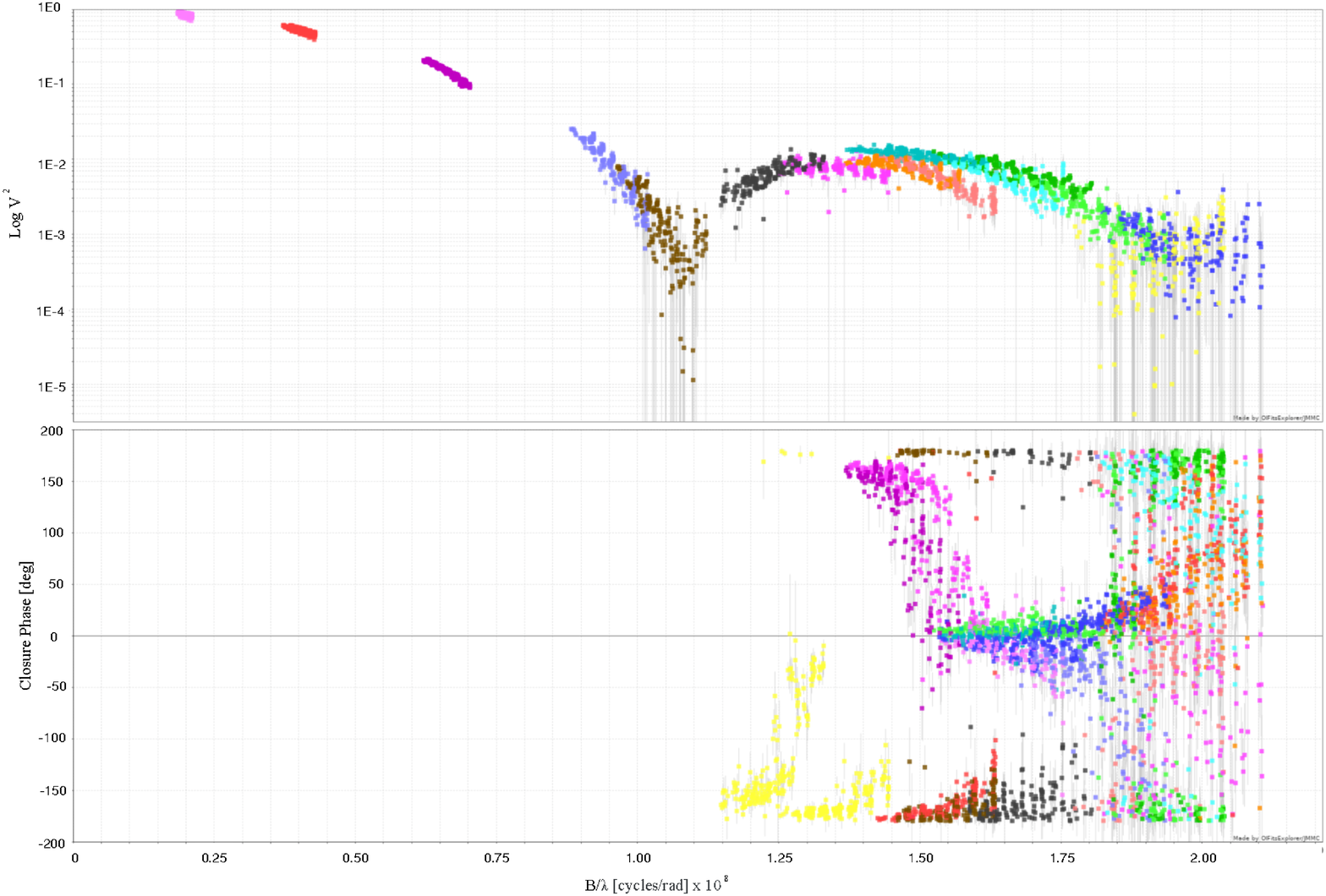}\\ 
        \includegraphics[width=0.95\hsize]{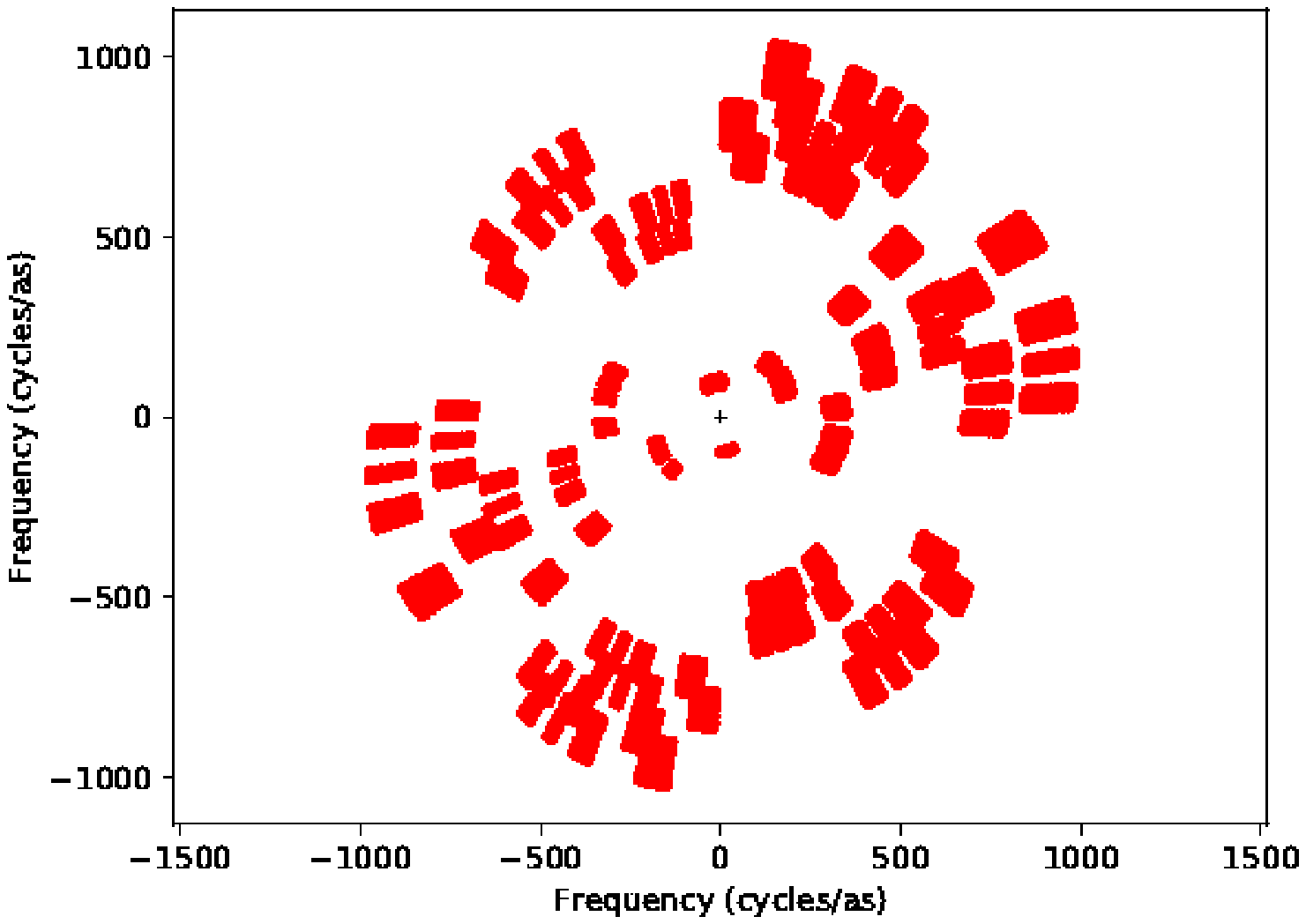} 
      \end{tabular}
      \caption{\emph{Top panels:} squared visibilities and closure phases for all the data of MIRC-X observations of CL~Lac. Visibilities are plotted in a log scale. The different colours indicate different baselines. \emph{Bottom panel:} UV coverage. The spectral coverage contains wavelengths between 1.52 and 1.70 $\mu$m. The resolution of the interferometer corresponds to a beam of $323.42\times323.42$ m telescope, $0.48\times0.48$ mas and $0.55\times0.55$ at 1.52 and 1.70 $\mu$m, respectively.}
        \label{observed_data}
\end{figure}   

 \begin{figure*}[!h]
   \centering
    \begin{tabular}{cc}
    \includegraphics[width=0.5\hsize]{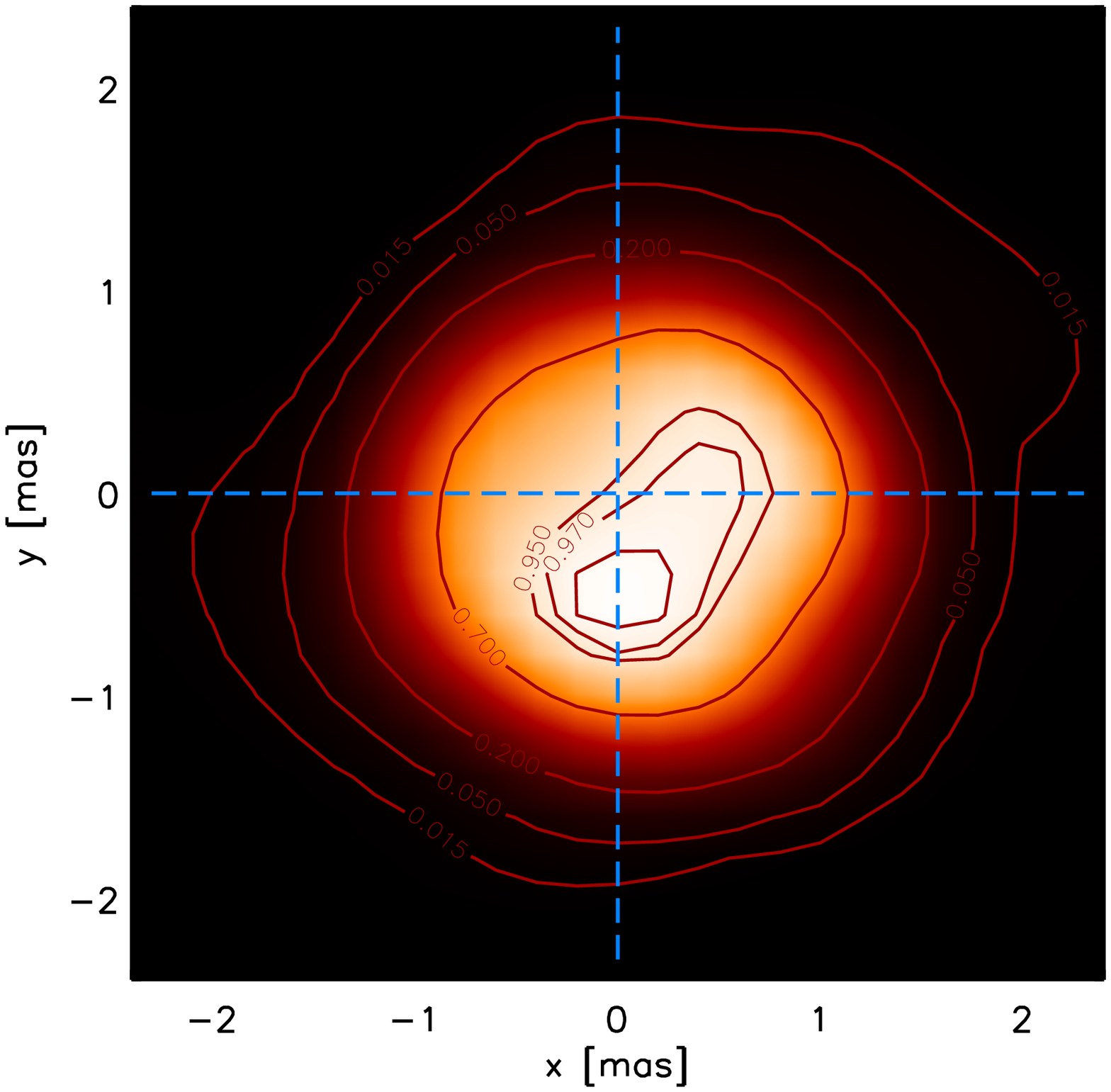} 
    \includegraphics[width=0.5\hsize]{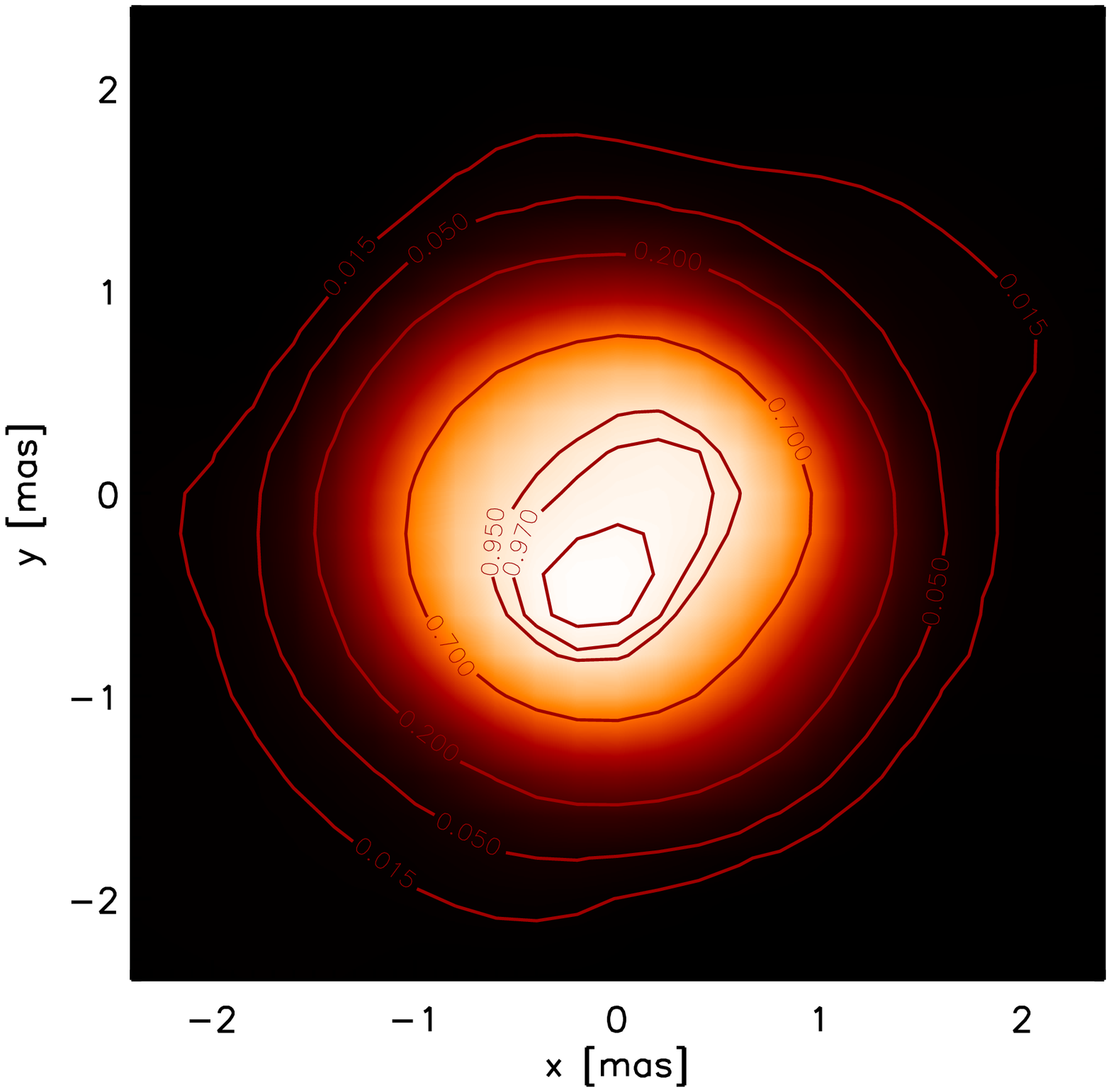} \\
    \includegraphics[width=0.5\hsize]{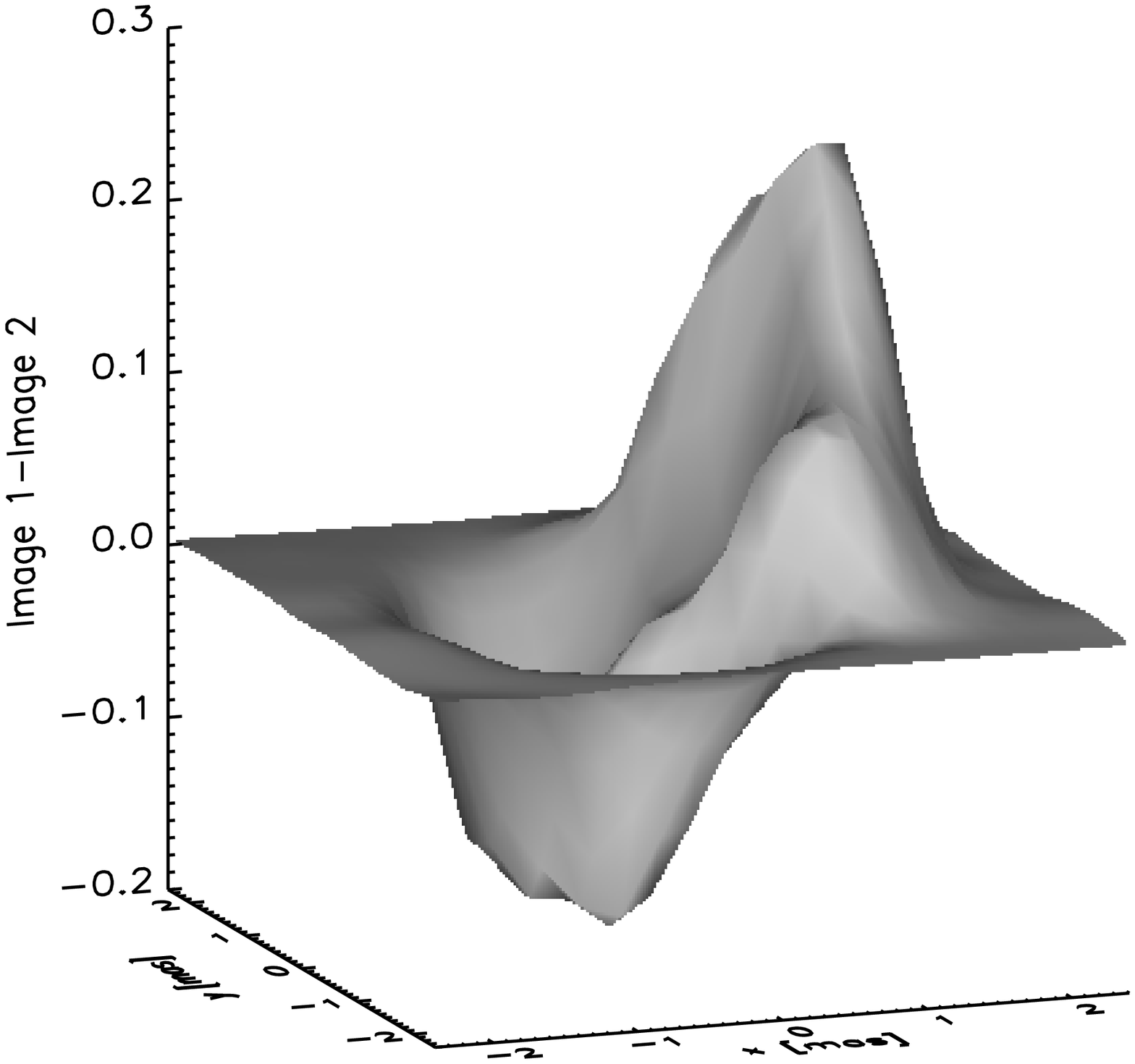} 
         \end{tabular}
      \caption{\emph{Top panels:} reconstructed images of CL~Lac at 1.52 $\mu$m (left) and 1.70 $\mu$m (right). They are convolved with the interferometric beam and the intensity is normalised between [0, 1]. The contour lines correspond to [0.015,0.05,0.20,0.70,0.95,0.97,0.99]. East is to the left, north is up. The barycentre of a symmetric star would be expected at x=0 and y=0, as shown by the blue dashed line. \emph{Bottom panel:} surface rendering difference between CL~Lac images at 1.52 and 1.70 $\mu$m.}
        \label{im1}
\end{figure*}   

The Gaia mission \citep{2016A&A...595A...1G} is an astrometric, photometric, and spectroscopic space-borne mission, which in 2018 delivered high-precision astrometric parameters (i.e. positions, parallaxes, and proper motions) for over one billion sources  \citep{2018A&A...616A...1G}. A considerable fraction of the intrinsically detected variable stars are long-period variables \citep[LPVs, of which the survey contains more than 150 000,][]{2018A&A...618A..30H} with large luminosity amplitudes and variability timescales, adequately sampled by Gaia \citep{2019A&A...623A.110G}. Among LPVs, there are cool luminous evolved stars with low to intermediate mass that reached a critical phase of their evolution at the end of the asymptotic giant branch (AGB). They are characterised by (i) large-amplitude variations in radius, brightness, and temperature of the star; (ii) a strong mass-loss rate \citep[$\dot{M}=10^{-6}$-10$^{-4}$ M$_\odot$/yr, ][]{2010A&A...523A..18D} driven by an interplay between pulsation, dust formation in the extended atmosphere, and radiation pressure on the dust \citep{2018A&ARv..26....1H}; (iii) an inhomogenous visible surface (produced in the interior and shaped by the top of the convection zone: they travel outwards on timescales ranging from weeks to years \citep{2017A&A...600A.137F}); and (iv) strong molecular absorption bands in the optically thin region \citep[e.g.][]{2000A&AS..146..217L} and on the top of the convection-related surface structures.

The concordance of all these properties makes the measurement of accurate stellar parameters and mass-loss rate very complex. In this context, a special emphasis has to be put on the problems with establishing their inter-dependence: a quantitative description of how the mass-loss rate depends on individual stellar parameters (and, consequently, how it changes as stars evolve) is essential for models of stellar and galactic chemical evolution \citep{2018A&ARv..26....1H}.

A noteworthy step forward in trying to find a way to quantitatively recover stellar parameters of AGB stars was proposed by \cite{2018A&A...617L...1C} using Gaia DR2 data and stellar convection simulations \citep{2017A&A...600A.137F}. The authors showed that the convection-related variability causes photocentre\footnote{The intensity-weighted mean of all emitting points tiling the visible stellar surface} displacements up to $\approx11\%$ of the stellar radius and accounts for a substantial part of the Gaia DR2 parallax error. They suggested that the fundamental properties of AGB stars could be measured directly from Gaia parallax errors. 
However, a final piece of information was missing, because the convection-related structures were quantified only indirectly (i.e. using photocentre displacement). Only with high angular resolution is it possible to detect the presence of stellar surface inhomogenities. This can be achieved using optical interferometry, which has already proven to be a powerful tool providing amazingly resolved aperture synthesis images of evolved stars \citep[e.g.][]{2010A&A...511A..51C,2014A&A...572A..17M,2016A&A...589A..91O,2016A&A...588A.130M,2017A&A...600L...2C,2017A&A...605A.108M,2017Natur.548..310O,2017A&A...606L...1W,2018A&A...614A..12M,2018Natur.553..310P}. 

In this work, we show the presence of inhomogeneities on the stellar surface of the AGB star CL~Lac, which is part of the sample of Gaia DR2 objects for which \cite{2018A&A...617L...1C} showed that photocentre displacement explains the parallax error bars. The article is structured as follows: Section~\ref{interfero} introduces the interferometric observations and data reduction, while Section~\ref{comparison} evidences the interpretation of the data with 3D simulation of stellar convection. Section~\ref{conclusions} summarises the conclusions and the impact on the determination of AGBs' stellar properties using Gaia parallax errors.

\section{Interferometric observations with MIRC-X at CHARA}\label{interfero}

The main objective of this work is to detect the presence of stellar surface inhomogeneities for an object from  Gaia's sample used in \cite{2018A&A...617L...1C}, and for which the convection-related variability accounts for a substantial part of the Gaia DR2 parallax error. We observed the LPV AGB star CL~Lac with the MIRC-X beam combiner \citep{2018SPIE10701E..23K,2018SPIE10701E..24A} and with the VEGA  instrument \citep{2009A&A...508.1073M} installed at 
Georgia State University Center for High Angular Resolution Astronomy (CHARA). The CHARA array is located on Mount Wilson, CA, and consists of six 1-m telescopes for a total of 15 baselines ranging in length from 34 m to 331 m, resulting in an angular resolution of 0.5 mas in the $H$ band \citep{2005ApJ...628..453T}. The observations took place on both UT 2019Jul27 and 2019Jul28. Some stellar properties of CL~Lac are reported in Table~\ref{CLLac_tab}.

\subsection{Data reduction}\label{sec_data_analysis}

We combined the light from all six telescopes to record three calibrated sets of data on each night on CL~Lac using MIRC-X in PRISM50 mode, which covers the $H$ band wavelength range  [1.52, 1.55, 1.58, 1.61, 1.64, 1.67, 1.70] $\mu$m. On each night, we recorded three sets of MIRC-X data on CL~Lac.  Each set consisted of 10-15 min to record fringe data and another 10 min to record backgrounds and shutter sequences.  In between these sets, we observed calibrator stars to calibrate the interferometric measurements.  The calibrators and adopted angular diameters are listed in Table~\ref{calibrators}. In order to verify, we calibrated the calibrators against each other on each night.  We saw no evidence for binarity (closure phases consistent with 0$^\circ$ within their uncertainties). A fit to the calibrator visibilities provided angular diameters consistent with the adopted values.

\begin{table*} 
\begin{center}
 \caption{Calibrators for CL~Lac}
 \label{calibrators}
 \begin{tabular}{ccccc}
\hline
  Name & Spectral type & Diameter [mas] & Observing night & Reference \\
\hline
HD$\_$210855 & F8V & $\Theta_\mathrm{UD}$=0.594$\pm$0.060 &  UT 2019Jul28 & JMMC, \cite{2017yCat.2346....0B} \\
HD$\_$219080 & F1V & $\Theta_\mathrm{LD}$=0.648$\pm$0.008  & UT 2019Jul27 and UT 2019Jul28 &  \cite{2013MNRAS.434.1321M} \\
HD$\_$571 & F5II& $\Theta_\mathrm{UD}$=0.609$\pm$0.059  & UT 2019Jul27 & JMMC, \cite{2017yCat.2346....0B}\\
\hline
 \end{tabular}
\end{center}
\end{table*}   

The data were reduced using the MIRC-X data reduction pipeline v1.2.0\footnote{https://gitlab.chara.gsu.edu/lebouquj/mircx$\_$pipeline.git.} to produce calibrated visibilities and closure phases. The MIRC-X detector is susceptible to bias in the bispectrum estimation as pointed out by \cite{2004MNRAS.347.1187B}. The data pipeline corrects for this bispectrum bias with a method similar to the one outlined in Appendix C of that paper.
 Figure~\ref{observed_data} displays the visibility and closure phase data for all the spectral channels as well as the UV-coverage. The data, and in particular the closure phases, denote the presence of large departure points from the centrosymmetric case (i.e. closure phases not equal to 0 or $\pm180^{\circ}$). The calibrated OIFITS files are available on the Optical interferometry DataBase\footnote{http://oidb.jmmc.fr}.

So as to obtain differential stellar radii in the optical and infrared, in order to retrieve information about the stellar dynamics and opacity run, we performed simultaneous  observations with VEGA. We used the MIRC-X beam combiner as a fringe tracker. The longitudinal dispersion correctors were used to compensate for differences in dispersion between the visible and near-infrared. However, no fringes were spotted out in the optical with VEGA. 
This can be due by the combination of a large object (> 4.5 mas) and low squared visibilities (<0.1), very close to the limit of the instrumental capabilities. Nevertheless, we cannot definitively conclude on this point because the SNR was significantly lower than the confidence limit of three.
   
\subsection{Aperture synthesis imaging}

For the image reconstruction, we used the MIRA software package developed by \cite{2008SPIE.7013E..1IT}. This package is written in the scientific language yorick\footnote{\url{https://software.llnl.gov/yorick-doc/}}. It reads interferometry data in the form of oifits files \citep{2005PASP..117.1255P}, and the reconstructed image is compared to the visibility and closure phase data by means of Fourier transforms and a few additional steps. The pixel values are modulated to minimise a so-called cost function, which is the sum of a regularisation term plus
data-related terms ($\chi^2$). The data terms enforce agreement of the model
image with the measured data (visibilities and closure phases).
The regularisation term is a distance minimisation between the
reconstructed image and an \emph{expected} image, called the prior. 

The data and the regularisation terms must be scaled to maximise the Bayesian evidence (the overall probability of the model integrated over all possible model parameter values). The scaling factor is called the 'hyperparameter' ($\mu$).

We reconstructed images for the spectral channels individually (Fig.~\ref{img_lambda}) with: 0.2 mas/pixel, a 64x64-pixel field of view (FOV), a total variation regularisation \citep{2011A&A...533A..64R}, a circular Gaussian prior image with a FWHM equal to half the field of view, and a hyperparameter value of $\mu=5\times10^4$. This value of $\mu$ is confirmed by the L-curve plot shown in Fig.~\ref{Lcurve_app}, which shows the data term ($\chi^2$) as a function of the hyperparameter ($\mu$, all other image reconstruction parameters are random: pixel size, number of pixels, prior size, number of iterations): one can see that the $\chi^2$ has a value close to 2 with $\mu \leq 50000$, and that the value raises by a large amount above. We show here the images convolved with a beam of FWHM $\lambda/2B$, meaning two times smaller as the theoretical angular resolution of the interferometer. This is to take advantage of the 'super-resolution' effect of image reconstruction with a sparse aperture reported in many works \citep[e.g. ][]{2008SPIE.7013E..1IT}. For convenience, the image convolved with a beam of FWHM $\lambda/B$ is also shown in Appendix B (Fig.~\ref{bis}).

We also reconstructed images using 32$\times$32 and 128$\times$128 pixels with the same pixel size (see Fig.~\ref{img_dep_FOV}), 64$\times$64 images with different pixel sizes (from 0.1 to 0.4 mas, Fig.~\ref{img_dep_px}), and hyperparameter values (from 5 to $5\times10^6$, see Fig.~\ref{img_dep_mu}), to ensure that the  images are not affected by reconstruction artifacts. We also checked that the global appearance of the image is similar when reconstructing a 'grey' image (i.e. computing an average across the spectral channels) and a chromatic image (Fig.~\ref{img_dep_FOV}).

Figure~\ref{im1} displays two examples of aperture synthesis images at 1.52 and 1.70 $\mu$m. They were convolved with the resolution of the interferometer (Fig.~\ref{observed_data}, bottom panel). Both images clearly show a brighter area in the south-eastern part of the map that affects the position of the barycentre, which does not correspond to the geometrical one (dashed blue line). The morphology of CL~Lac is complex and not centrosymmetric, largely affecting astrometric Gaia measurements \citep{2018A&A...617L...1C}. These stellar surface inhomogenities show a similar behaviour for all the spectral channels but with noticeable changes between 1.52 and 1.70 $\mu$m. Figure~\ref{im1} (bottom panel) displays large differences (up to $\sim20\%$), in particular close to the outer stellar disc borders.

\section{Interpreting stellar surface inhomogeneities}\label{comparison}

\subsection{Three-dimensional simulations of AGB stars and synthetic images}

We use the radiation-hydrodynamics (RHD) simulation of AGB stars \citep{2017A&A...600A.137F} computed with CO$^5$BO1D \citep{2012JCoPh.231..919F} code. The code solves the coupled non-linear equations of compressible hydrodynamics and non-local radiative energy transfer in the presence of a fixed external spherically symmetric gravitational field on a 3D cartesian grid. The averaged stellar parameters of the simulation used in this work are reported in Table~\ref{simu}. The atmosphere in the simulation is characterised by global and small-scale shocks induced by large-amplitude, radial, and fundamental-mode pulsations \citep{2017A&A...600A.137F}. The pulsation period is extracted with a fit of the Gaussian distribution in the power spectra of the simulations. The shocks contribute to the levitation of material and to the photocentre displacement \citep[Fig.~A.5 for the temporal behaviour, ][]{2018A&A...617L...1C}. A photocentre displacement of $\sigma_P=0.131$ AU (Table~\ref{simu}) corresponds to $\approx27 R_\sun$ or $\approx$9$\%$ of the stellar radius.

To prepare the comparison with the interferometric data, we computed intensity maps in the same observed spectral channels (Fig.~\ref{im1}) for the 100 snapshots of the RHD simulation. %Each map being an average of 5 intensity maps, which correspond to 5 different wavelength points (i.e., a computational resolving power of $\lambda/\Delta\lambda=$20000).
For this purpose, we employed the code
{{\sc Optim3D}} \citep{2009A&A...506.1351C}, which takes into account the Doppler shifts caused by the
convective motions. The radiative transfer is computed in detail using pre-tabulated extinction coefficients per unit mass, like for MARCS (\citealp{2008A&A...486..951G}) as a function of temperature, density, and wavelength for the solar composition \citep{2009ARA&A..47..481A}. The
temperature and density distributions are optimised to cover the values
encountered in the outer layers of the RHD simulation.

 \begin{figure}
   \centering
    \begin{tabular}{c}
    \includegraphics[width=1.\hsize]{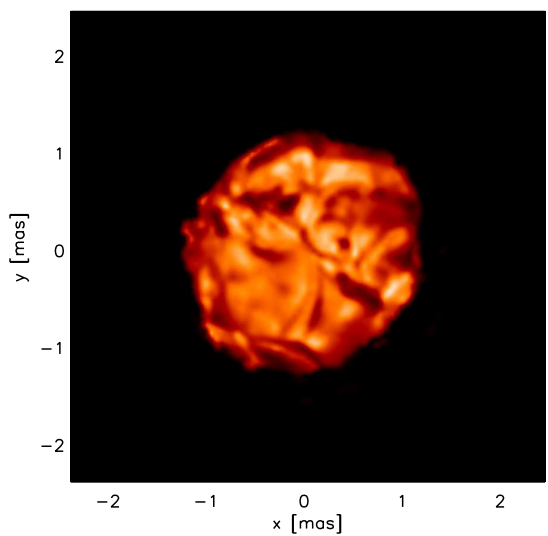} \\
      \end{tabular}
      \caption{Synthetic intensity map computed for one snapshot of the RHD simulation of Table~\ref{simu} at 1.52 $\mu$m. The intensity is normalised between [0, 1]. East is left, north is up. This image is the non-convolved version of  Fig.~\ref{3D_images} (top-left panel).}
        \label{3D_images_example}
\end{figure} 

\begin{table*}
\footnotesize 
\begin{center}
 \caption{Parameters of the RHD simulation used in this work. This simulation is part of the AGB grid of \cite{2017A&A...600A.137F}}
 \label{simu}
 \begin{tabular}{l|ccccccc|cc}
\hline
Simulation & $M_\star$\tablefootmark{a} & $L_\star$\tablefootmark{a} & $R_\star$\tablefootmark{a} & $T_\mathrm{eff}$\tablefootmark{a} & $\log g$\tablefootmark{a} & $t_\mathrm{avg}$\tablefootmark{a} & $P_\mathrm{puls}$\tablefootmark{a} & $P$\tablefootmark{b} & $\sigma_P$\tablefootmark{b}  \\
& $M_\sun$ & $L_\sun$  & $R_\sun$ & K & (cgs) & yr & yr &  AU & AU \\
\hline
st28gm05n001 & 1.0 & 5040 $\pm$195 & 306 $\pm$18 & 2790$\pm$59 & -0.54$\pm$0.05 & 25.36 & 1.026$\pm$0.135 & 0.183 & 0.131 \\
\hline
 \end{tabular}
\end{center}
\tablefoot{The first column shows the simulation name, then the next five columns the stellar parameters such as the mass ($M_\star$), the average luminosity ($L_\star$), radius ($R_\star$), effective temperature 
($T_\mathrm{eff}$), and surface gravity ($\log g$). The seventh column displays the stellar simulated time ($t_\mathrm{avg}$) used to average all the quantities. A solar metallicity is assumed. The eighth and ninth columns report the pulsation period ($P_\mathrm{puls}$) and the half of the distribution of the pulsation frequencies ($\sigma_{\mathrm{puls}}$). Finally, the last two columns are the time-averaged value of the photocentre displacement $P$ and its standard deviation ($\sigma_P$).\\
\tablefoottext{a}{\cite{2017A&A...600A.137F}}\\
\tablefoottext{b}{\cite{2018A&A...617L...1C}}
}
\end{table*}

\subsection{Stellar radius thanks to the Gaia parallax}\label{sect_radius}

 \begin{figure}
   \centering
    \begin{tabular}{c}
        \includegraphics[width=0.98\hsize]{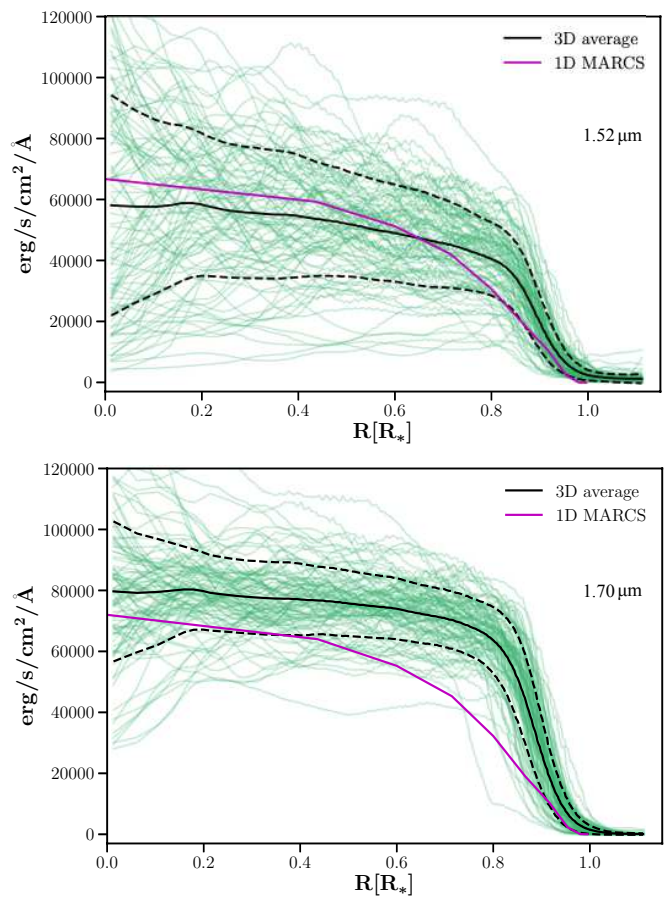}
      \end{tabular}
      \caption{Azimuthally averaged intensity profiles (green curves) obtained from the intensity map of each simulation snapshot of Table~\ref{simu}. The black curve is the temporally averaged intensity profile, while the dashed black lines denote the 1-$\sigma$ temporal fluctuations. The purple line in the 1D intensity profile from the best-fitting MARCS model (see text).}
        \label{radially_averaged}
\end{figure}   

\begin{table}[!h]       
\footnotesize
\begin{center}
 \caption{CL~Lac apparent diameters obtained with the best-fitting 1D-MARCS model, $\Theta_\mathrm{1D}$, and the one from the 3D temporally averaged intensity profile (Fig.~\ref{radially_averaged}, black curve), $\Theta_\mathrm{3D}$.}
 \label{diameters}
 \begin{tabular}{ccccc}
\hline
  Wavelength & $\Theta_\mathrm{1D}$ & $R_\star$ &  $\Theta_\mathrm{3D}$ & $R_\star$ \\
  $\mu$  & mas & $R_\sun$ & mas & $R_\sun$\\
\hline
1.52  & 3.332$\pm$0.005 & 310$\pm$41 & 3.299 $\pm$0.005 &  307$\pm$41 \\
1.70 & 3.220$\pm$0.004 & 300$\pm$40 & 3.053 $\pm$ 0.006 &  284$\pm$38 \\
\hline
 \end{tabular}
\end{center}
\end{table}   

The first step of our analysis is the determination of the stellar radius. In order to do this, we used (i) 1D, stationary, hydrostatic models; and (ii) the 3D RHD simulation of Table~\ref{simu}. 

 For  item (i), we selected several models from the MARCS-grid\footnote{http://marcs.astro.uu.se} \citep{2008A&A...486..951G} with stellar parameters close to CL~Lac (Table~\ref{CLLac_tab}): effective temperature $T_\mathrm{eff}$=[2500, 2600, 2700, 2800], mass=1 $M_\sun$, $\log g$=[-0.5, 0.0], metallicity [Fe/H] = [$0.0$]. We used the code TurboSpectrum \citep{2012ascl.soft05004P} to compute intensity profiles from these models in the same observed spectral channels of Fig.~\ref{im1}.\\
  For item (ii), we obtained the 100 azimuthally averaged intensity profiles from the synthetic maps using the method explained by \cite{2009A&A...506.1351C}. They were constructed using rings that are regularly spaced in $\mu$ (where $\mu = \cos(\theta)$, with $\theta$ being the angle between the line of sight and the radial direction). Figure~\ref{radially_averaged} shows that the temporal fluctuation from one snapshot to another (green curves) is large, each snapshot being approximatively 23 days apart. 
        Afterwards, we calculated synthetic visibility amplitudes from the different 1D-MARCS models (purple curve) and the 3D temporally averaged (black curve) intensity profiles using the Hankel transform. Finally, we fitted the visibility data from Fig.~\ref{observed_data} separately for the two spectral channels at 1.52 and 1.70 $\mu$m to obtain the apparent diameter of CL~Lac. The best-fitting MARCS model has $T_\mathrm{eff}$=2800K and $\log g$=-0.5. \\
        Table~\ref{diameters} reports the 1D and 3D stellar diameters as well as the stellar radius at a distance of $d=870\pm113$ pc (i.e. inversion of the Gaia parallax, Table~\ref{CLLac_tab}). The 1D-MARCS model returns apparent radii larger than the 3D, $\sim1\%$ and $5\%$ at 1.52 and 1.70 $\mu$m, respectively. This is already noticeable from the intensity profiles in Fig.~\ref{radially_averaged}. Our result is in qualitative agreement with those obtained by several authors for dwarf and K~giant stars \citep{2002ApJ...567..544A, 2005ApJ...633..424A, 2006A&A...446..635B, 2010A&A...524A..93C, 2012A&A...545A..17C, 2017A&A...597A.137K}. Moreover, wavelength-dependence of the radius is noticeable: this is a signature of different contributions of molecular absorption \citep{2016A&A...587A..12W}, even if the interpretation remains difficult at the actual resolution of our observations.
                        
        To sum up, the RHD simulation used here is the closest available to CL~Lac in terms of
\begin{enumerate}
    \item stellar parameters with, in particular, a very similar radius ($R_\star$=306$\pm$18, Table~\ref{simu}) with respect to the value we found (Table~\ref{diameters}) and pulsation period;
    \item photocentre variability in the Gaia G~band ($\sigma_P$=0.131, Table~\ref{simu}) caused by convection-related structures that accounts, as demonstrated by \cite{2018A&A...617L...1C}, for a substantial part of the CL~Lac parallax error ($\sigma_\varpi$=0.130, Table~\ref{CLLac_tab}).
\end{enumerate}

It has to be noted that the $T_\mathrm{eff}$ based on Gaia Bp/Rp photometric colours from Table~\ref{CLLac_tab} is almost 500K larger than the 1D MARCS and 3D RHD simulation, that, in turn, have overlapping values within error bars. However,  due to the limited training set for cool stars in Gaia actual DR2, photometric temperatures have to be taken with extreme caution for these stars.
        
        \subsection{Comparison with the 3D RHD simulation}

In this section, we compare the synthetic maps to the aperture synthesis imaging. As a first step, we correct the limb-darkening (LD) effect dividing both the synthetic and reconstructed images. We proceeded as follows: (i) we created an LD model image (64$\times$64 pixels and 0.2 mas/pixel resolution) with the same FOV as the reconstructed image using the 3D temporally averaged profile of Fig.~\ref{radially_averaged}; (ii) we rebinned the resolution of 0.2 mas/pixel the 3D synthetic maps so that they effectively possess the same FOV and resolution as the reconstructed images; and (iii) we convolved them to the resolution of the interferometric observations (Fig.~\ref{observed_data}, bottom panel) for both spectral channels at 1.52 and 1.70 $\mu$m.

To have strong constraints on the RHD simulation, we simultaneously inspected three different quantities for each map:
[1] the intensity contrast in the maps;
[2] the morphology of the surface structures; and [3] the position of the photocentre.
We performed a minimisation procedure for each item to obtain the normalised $\chi^2_\mathrm{compiled}=\chi^2_\mathrm{contrast}+\chi^2_\mathrm{morphology}+\chi^2_\mathrm{photocentre}$. 

\begin{figure}
   \centering
    \begin{tabular}{c}
     \includegraphics[width=1.0\hsize]{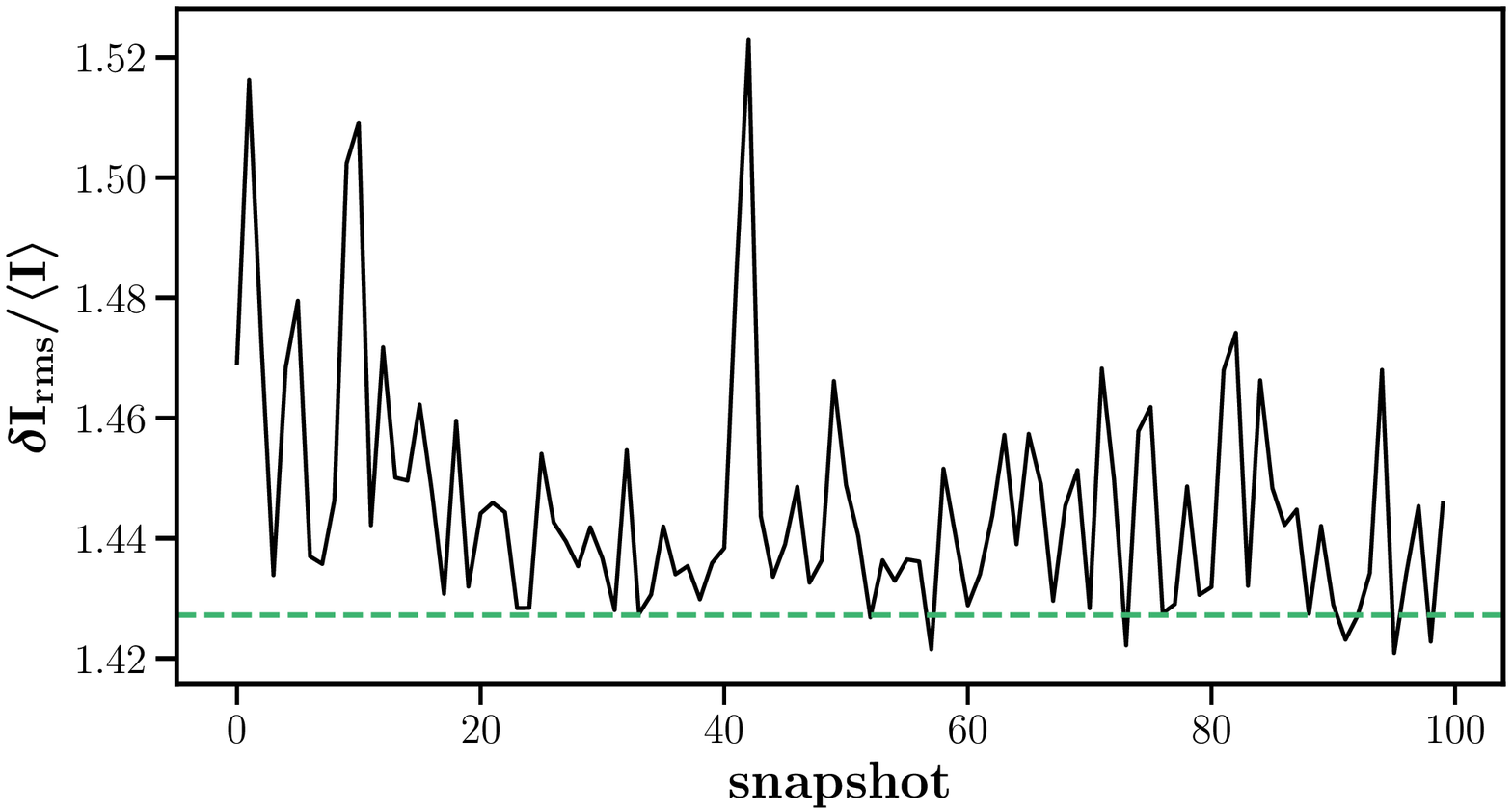} \\
     \includegraphics[width=1.0\hsize]{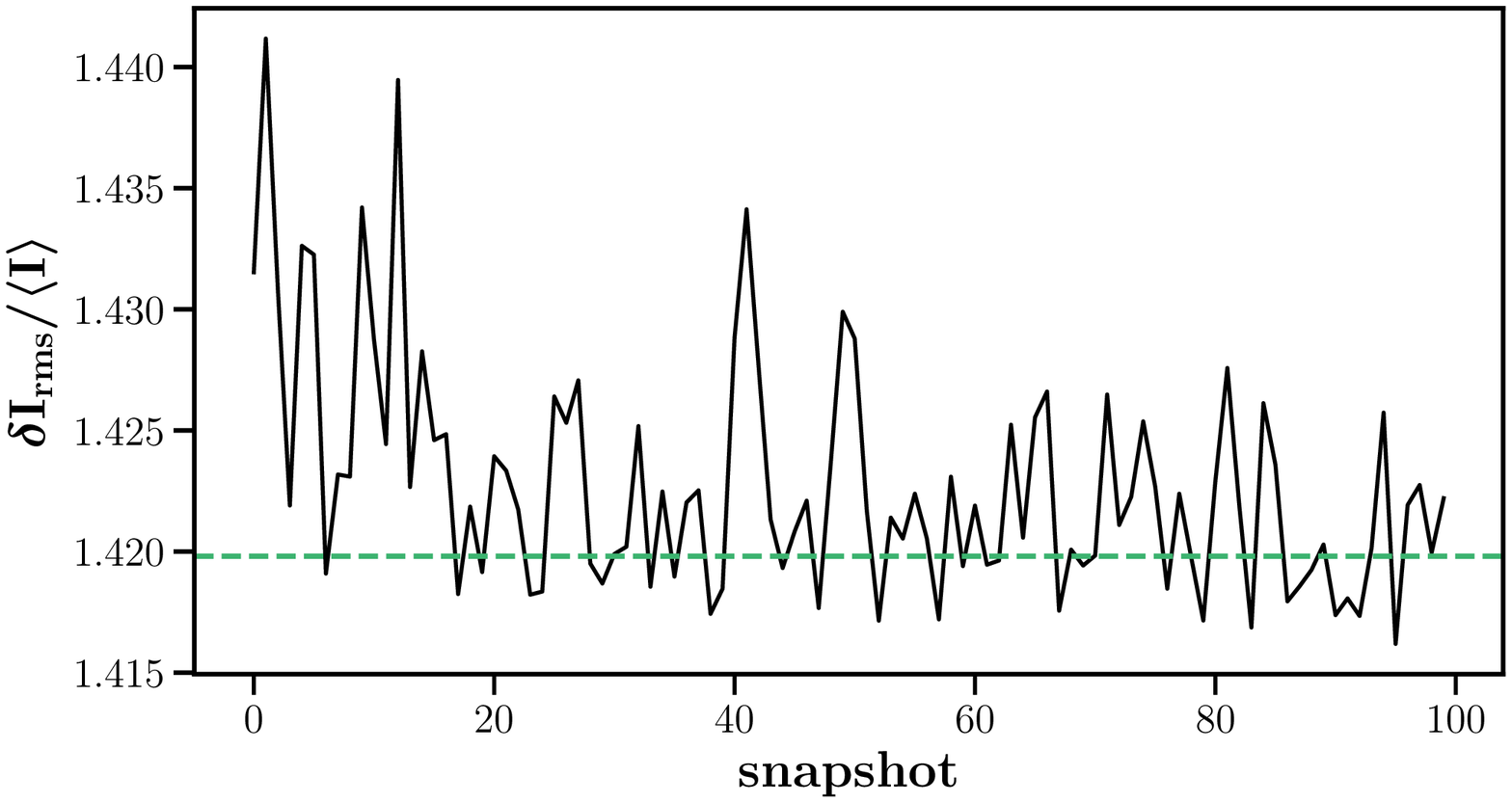} 
      \end{tabular}
      \caption{Fluctuations at 1.52 $\mu$m (top) and 1.70 $\mu$m (bottom) for all the snapshots of the RHD simulation of Table~\ref{simu} after being corrected for limb-darkening and convolved with the interferometric beam. The horizontal dashed line displays the contrasts measured in the reconstructed images of Fig.~\ref{im1}: 1.424 at 1.52 $\mu$m and 1.423 at 1.70 $\mu$m.}
        \label{fluctuations}
\end{figure}  

Starting from point [1], we divided pixel by pixel reconstructed images, and we convolved 3D synthetic maps by the LD one and used the definition of \cite{2013A&A...557A...7T} to estimate the intensity contrast as $\delta I_{\mathrm{rms}}/\langle I\rangle$. 
The contrast depends on the radial cut chosen \citep[i.e. the maximum radius adopted from the centre of the star from which outer pixels are not considered,][]{2018A&A...614A..12M}. For CL~Lac, we chose to cut at 0.80 stellar radii to avoid outer areas with rapid and steep increase. Figure~\ref{fluctuations} shows that several snapshots have very close values of contrast with respect to observations for both wavelength channels. The best matching 3D synthetic snapshot is the one with the closest intensity contrast with respect to the reconstructed images.\\
Furthermore, we concentrated our analysis on the morphology of the surface structures (point [2]). To account for the unknown orientation of CL~Lac with respect to the synthetic maps in the sky, we rotated the latter every 5$^\circ$ around its centre, and we calculated the minimum difference with the reconstructed images. Several snapshots were found to be very similar to the observed images, but, as expected, none matched perfectly. \\
Finally, we computed the position of the photocentre (point [3]) for each map and for the reconstructed images as the intensity-weighted mean of the $x-y$ positions of all emitting points tiling the visible stellar surface \citep[Eq.~1 and 2 from][]{2018A&A...617L...1C}. Then, we found the smallest difference between the photocentre in the observations and the synthetic maps. Among the different contributors to the $\chi^2_\mathrm{compiled}$, the morphology of the surface structures ($\chi^2_\mathrm{morphology}$) is the most dominant.

\begin{figure}
   \centering
    \begin{tabular}{cc}
     \includegraphics[width=0.9\hsize]{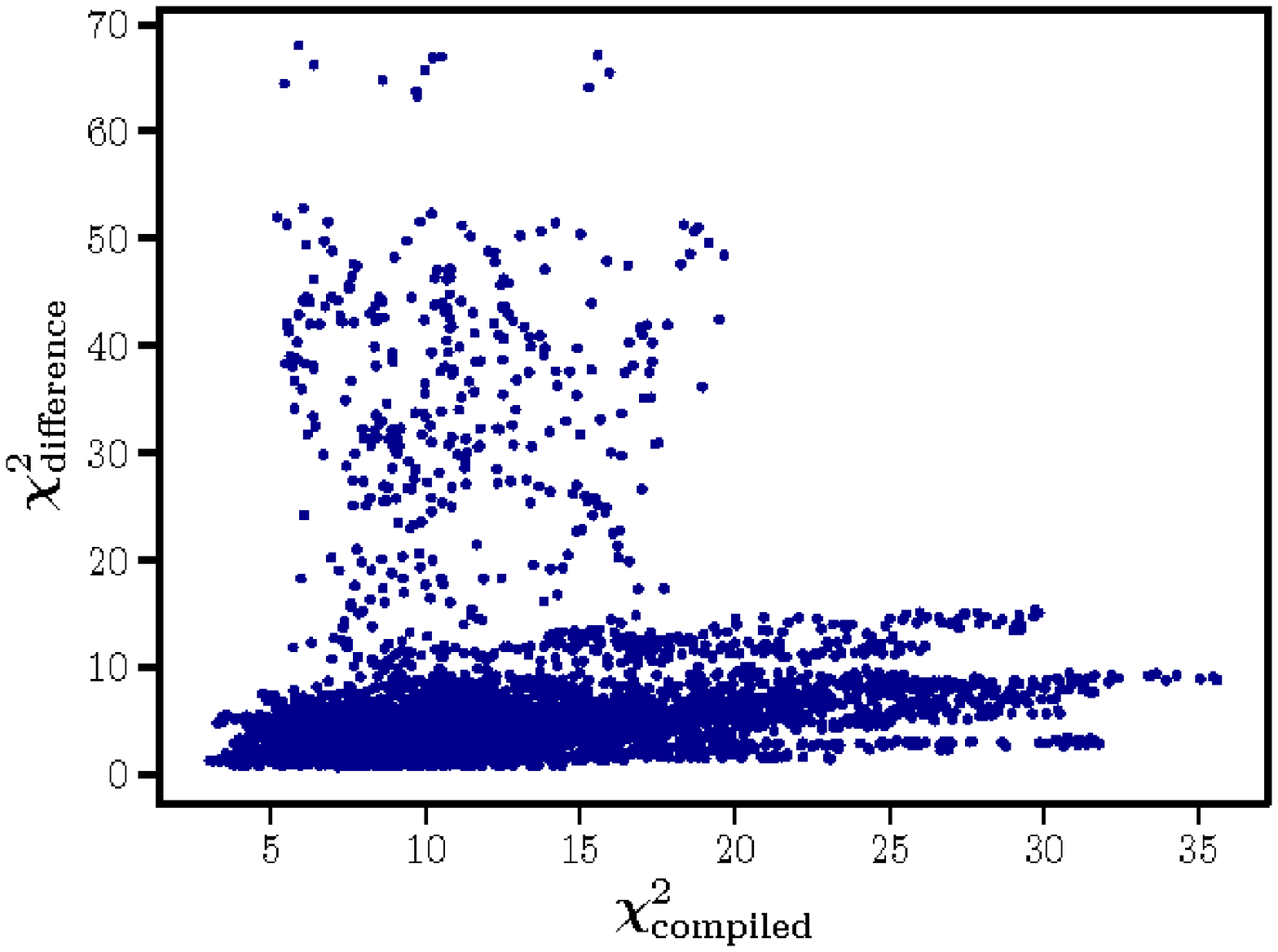} \\
    \includegraphics[width=0.9\hsize]{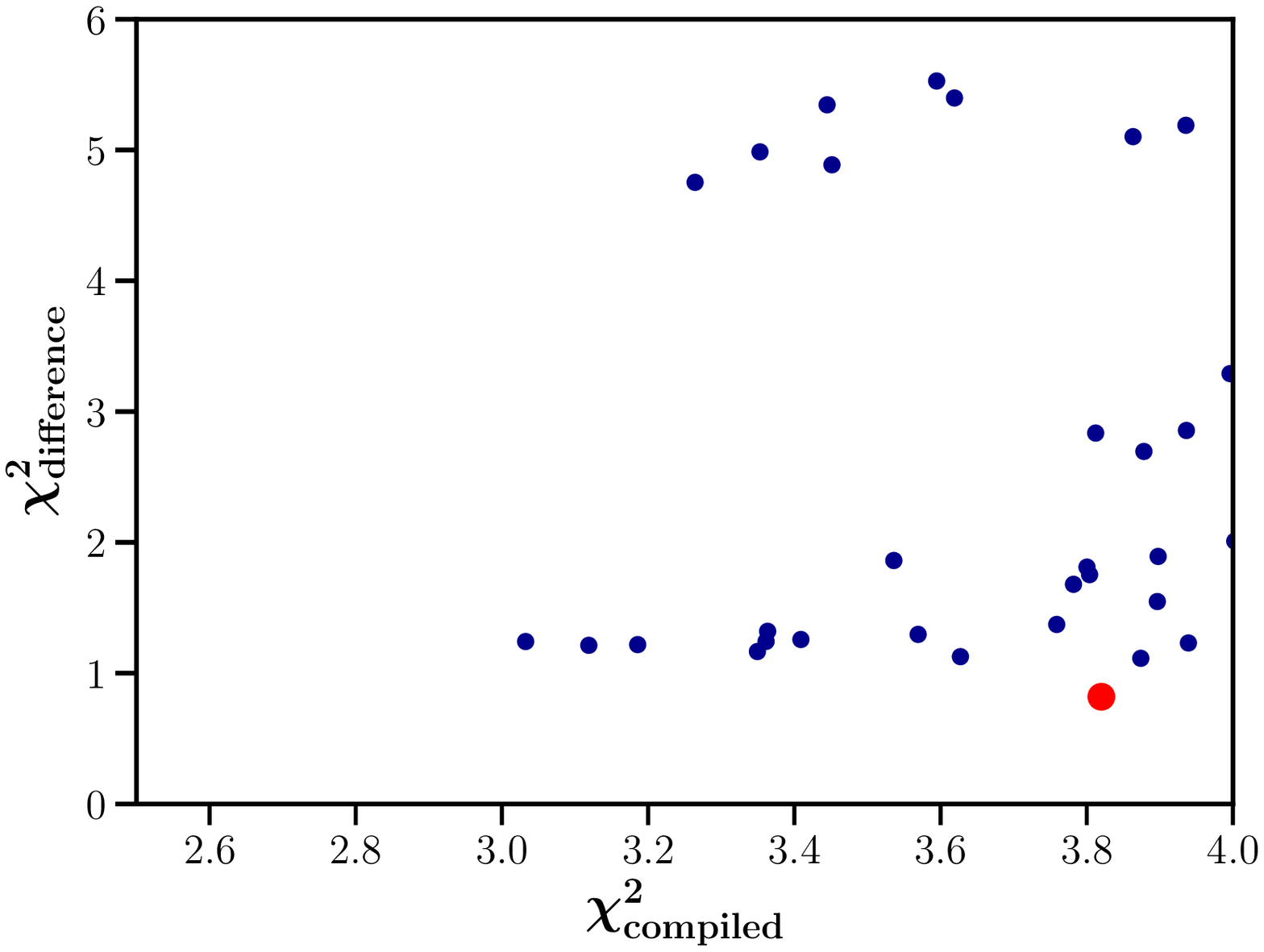} 
      \end{tabular}
      \caption{ \emph{Top panel:} compiled $\chi^2_\mathrm{compiled}=\chi^2_\mathrm{contrast}+\chi^2_\mathrm{morphology}+\chi^2_\mathrm{photocentre}$ versus $\chi^2_\mathrm{difference}=\Delta_{reconstructed}-\Delta_{3D}$ (see text for details) for all the snapshots and rotation angles considered (blue points) for the comparison with the reconstructed images. \emph{Bottom panel:} enlargement within the region where the best matching snapshot and rotation angle is chosen (large red point).}
        \label{chisquare}
\end{figure}

\begin{figure*}
   \centering
    \begin{tabular}{cc}
     \includegraphics[width=0.5\hsize]{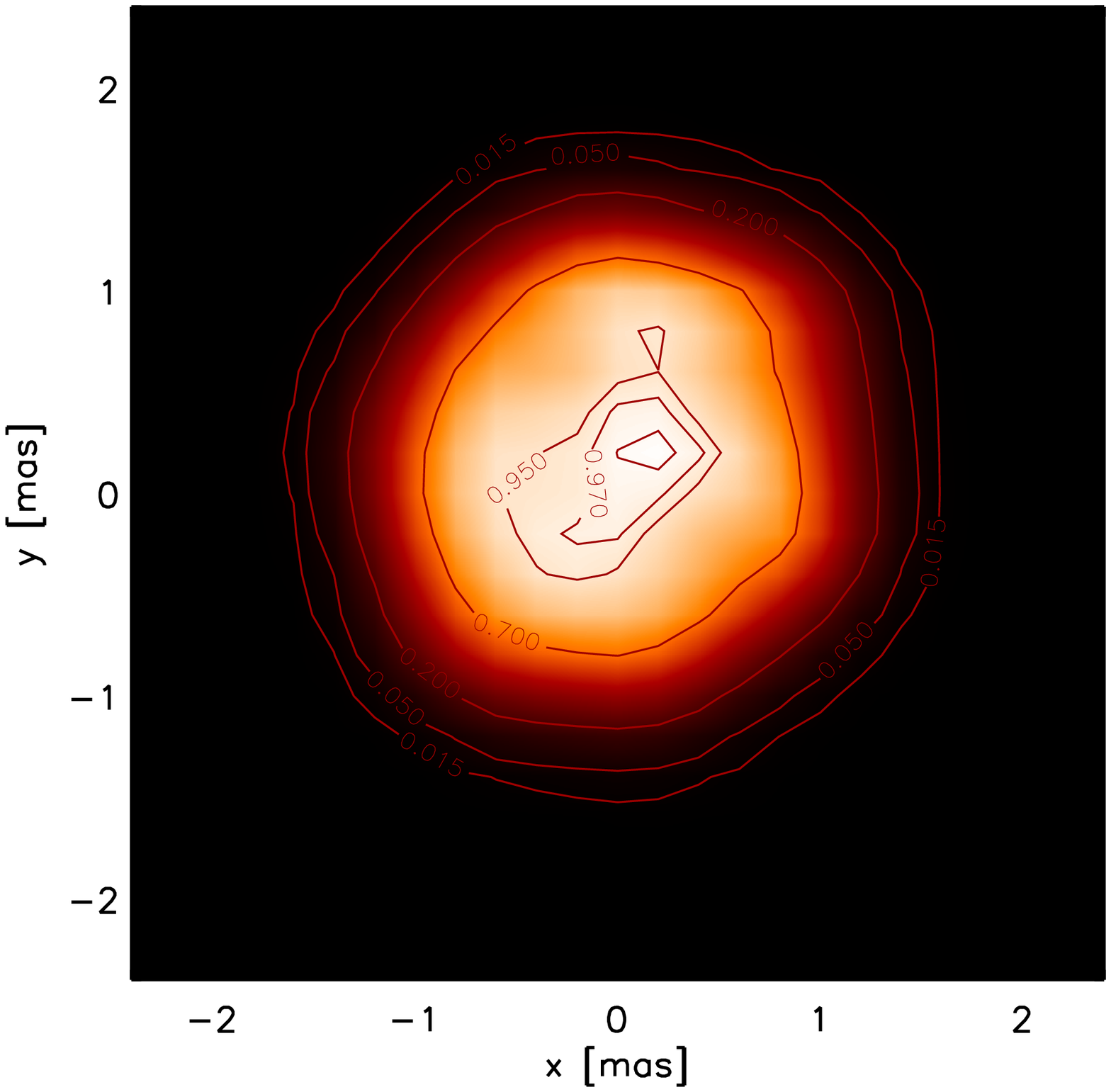} 
    \includegraphics[width=0.5\hsize]{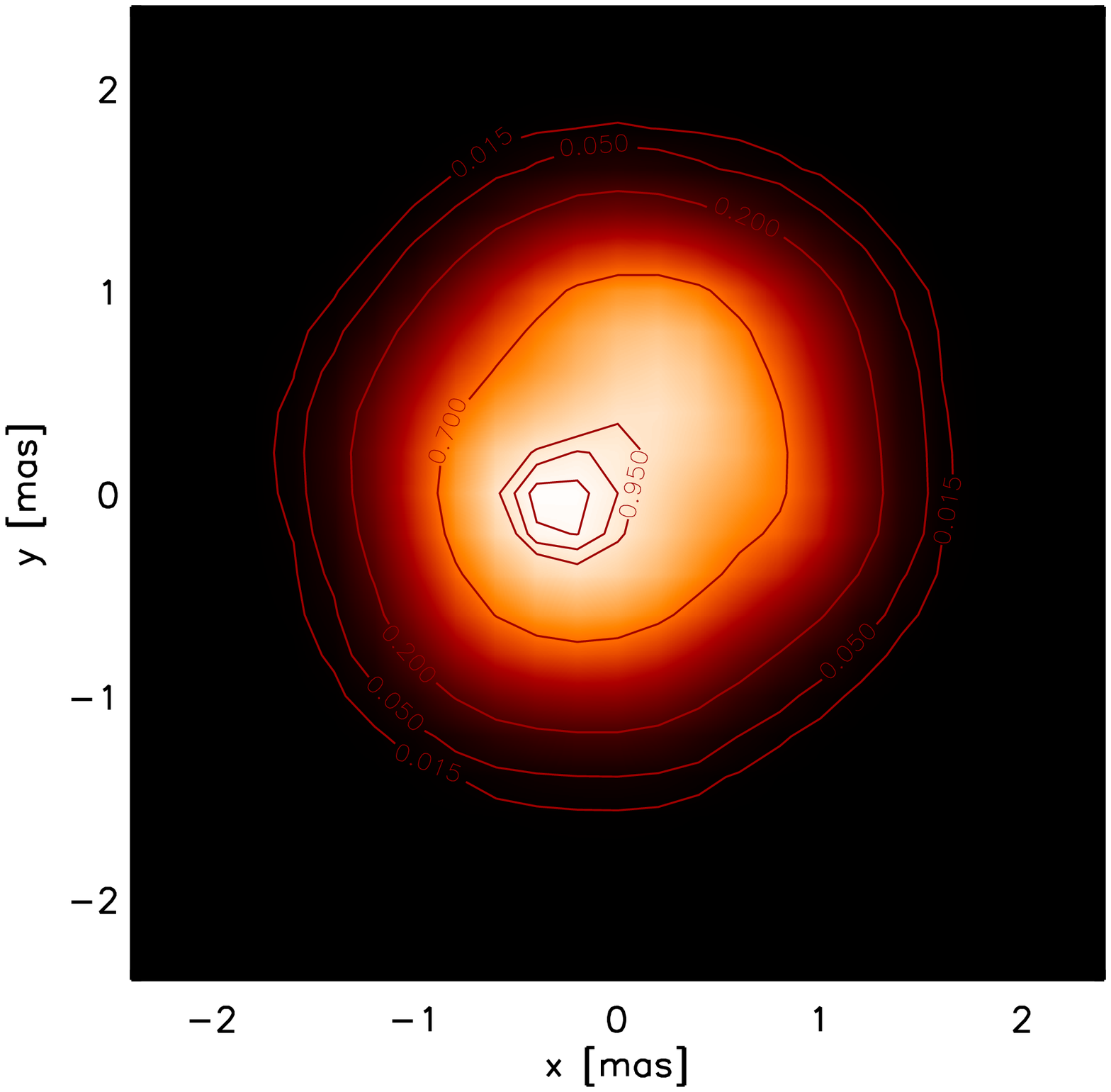} \\
    \includegraphics[width=0.5\hsize]{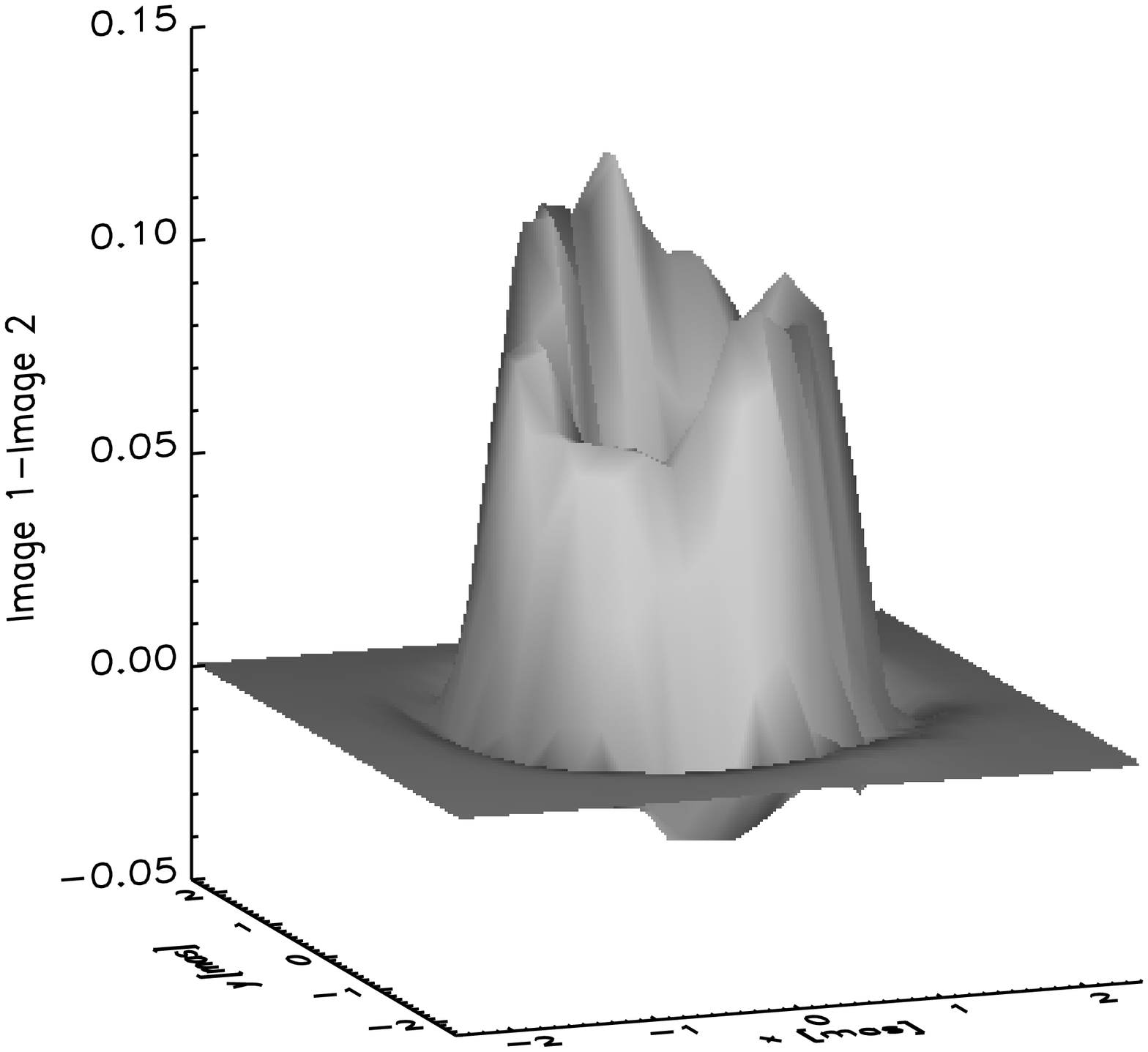} 
      \end{tabular}
      \caption{\emph{Top panels:} synthetic intensity maps computed for the best matching RHD simulation of Table~\ref{simu} at 1.52 (left) and 1.70 (rigth) $\mu$m. The intensity is normalised between [0, 1]. The images, 64$\times$64 pixels with 0.2 mas/pixel, were convolved with the interferometric beam (Fig.~\ref{observed_data}). The contour lines are the same as in Fig.~\ref{im1}. East is left, while north is up. \emph{Bottom panel:} surface rendering difference between the synthetic images at 1.52 and 1.70 $\mu$m (right).}
        \label{3D_images}
\end{figure*}

    In addition to these three points, we also considered the difference between the reconstructed images at 1.52 and 1.70 $\mu$m ($\Delta_{reconstructed}=image_{1.52}-image_{1.70}$, bottom panel of Fig.~\ref{im1}) and for the simulation images ($\Delta_{3D}=image3D_{1.52}-image3D_{1.70}$, bottom panel of Fig.~\ref{3D_images}). We aimed to have the closest difference ($\Delta$) between the two cases, and we defined $\chi^2_\mathrm{difference}=\Delta_{reconstructed}-\Delta_{3D}$. This approach gives an additional and important wavelength dependence constraint for the simulation. In the end, the best matching snapshot is the one with the combination of the smallest $\chi^2_\mathrm{compiled}$ versus $\chi^2_\mathrm{difference}$, with a higher weight for the latter. Figure~\ref{chisquare} shows all the snapshots and all the different rotation angles considered, while Fig.~\ref{3D_images} displays the best matching snapshot and rotation angle: the best combination returns $\chi^2_\mathrm{compiled}=$3.8 and $\chi^2_\mathrm{difference}=$0.8.

Our 3D RHD simulation shows a good agreement with the observations both in terms of contrast an surface structures and wavelength dependence, meaning that it is adequate for explaining the inhomogenities observed in the reconstructed maps. This definitively confirms the presence of  convection-related surface structures on the same Gaia DR2 object for which \cite{2018A&A...617L...1C} proved that photocentre displacement explains the parallax error bars.

\section{Discussion and conclusions}\label{conclusions}

 Stellar dynamics induce an intrinsic noise that increases the measurement uncertainty of the parallax of AGB stars in Gaia DR2 \citep{2018A&A...617L...1C}, among which there is the star CL~Lac.  However, the effect of convection-related surface structures was estimated only indirectly using the displacement of the photocentre ($\sigma_P$). In addition to this, the authors showed also a clear correlation between stellar parameters (surface gravity and pulsation period) and the standard deviation of the photocentre movements: 3D simulations with a lower surface gravity (i.e. more extended atmospheres) return larger excursions of the photocentre.
 
 Thanks to optical interferometry, we managed, in this work, to discover the presence of stellar inhomogenities on the AGB CL~Lac. Its morphology is complex, not centrosymmetric, and likely affects astrometric Gaia measurements. Using state-of-the-art 3D global simulations of stellar convection with CO$^5$BOLD, we were able to interpret them as a  consequence of small- and global-scale shocks that contribute to the levitation of material. This definitively confirms the presence of  convection-related surface structures on the same Gaia DR2 object for which \cite{2018A&A...617L...1C} discussed the photocentre displacement. 
 
 Our result will help us to make a step forward in exploiting Gaia measurement uncertainties using appropriate RHD simulations. The long-term aim is to uniquely extract the fundamental properties of AGB stars directly from Gaia errors and make it possible to systematically study the properties of convection in stars other than the Sun. These properties are paramount to understanding the physics of these stars: for example, the surface gravity controls the size of granules that, in turn, contribute to the stellar photometric variations. Similarly, the pulsation period provides important information about stellar (mean) interior as well as global shocks induced by large-amplitude, radial, and fundamental-mode pulsations. \\
 To achieve our goal, a series of steps are required. We list them here:
\begin{enumerate}
    \item An additional study of the photometric colours from Bp and Rp photometric systems; \cite{2011A&A...528A.120C} showed that red supergiant stars\footnote{Massive \citep[$M>8M_\sun$][]{2018A&ARv..26....1H}, cool evolved stars with stellar parameters close to AGBs.} have magnitude excursions up to 0.28 and 0.13 magnitudes in Bp and Rp, respectively. This impacts the determination of their parameters. We expected similar (or higher) values for AGBs.
    \item One limitation of the existing model grid \citep{2017A&A...600A.137F} is the restriction to 1 $M_\odot$.  A significantly larger RHD simulation grid with a larger range of stellar masses is necessary to explore the large number of AGBs in Gaia \citep{2018A&A...618A..30H}.
    \item With Gaia DR2, the mean number of measurements for each source amounts to $\approx$ 26 \citep{2018A&A...618A..58M}. Finally, when Gaia DR4 is available in 2022, the number of measurements will increase to 70-80, possibly changing the parallax error. Additionally, DR4 will provide time-dependant measurements allowing a direct and more precise comparison with the time-dependent RHD simulations.
\end{enumerate}

%--------------------------------------------------------------



\begin{acknowledgements}
This work is based upon observations obtained with the Georgia State University Center for High Angular Resolution Astronomy Array at Mount Wilson Observatory.  The CHARA Array is supported by the National Science Foundation under Grant No. AST-1636624 and AST-1715788.  Institutional support has been provided from the GSU College of Arts and Sciences and the GSU Office of the Vice President for Research and Economic Development. MIRC-X received funding from the European Research Council (ERC) under the European Union's Horizon 2020 research and innovation programme (Grant No. 639889). We also thank the grants NSF 1506540 and NASANNX16AD43G.\\
This work has made use of the MIRA software, kindly developed by E.~Thi\'ebaut who we thank for his great contribution to the community.
\end{acknowledgements}

\bibliographystyle{aa}
\bibliography{biblio.bib}

%--------------------------------------------------------------

\begin{appendix} %First appendix

\section{Reconstructed images for individual spectral channels}

 \begin{figure*}
   \centering
    \includegraphics[height=0.5\hsize]{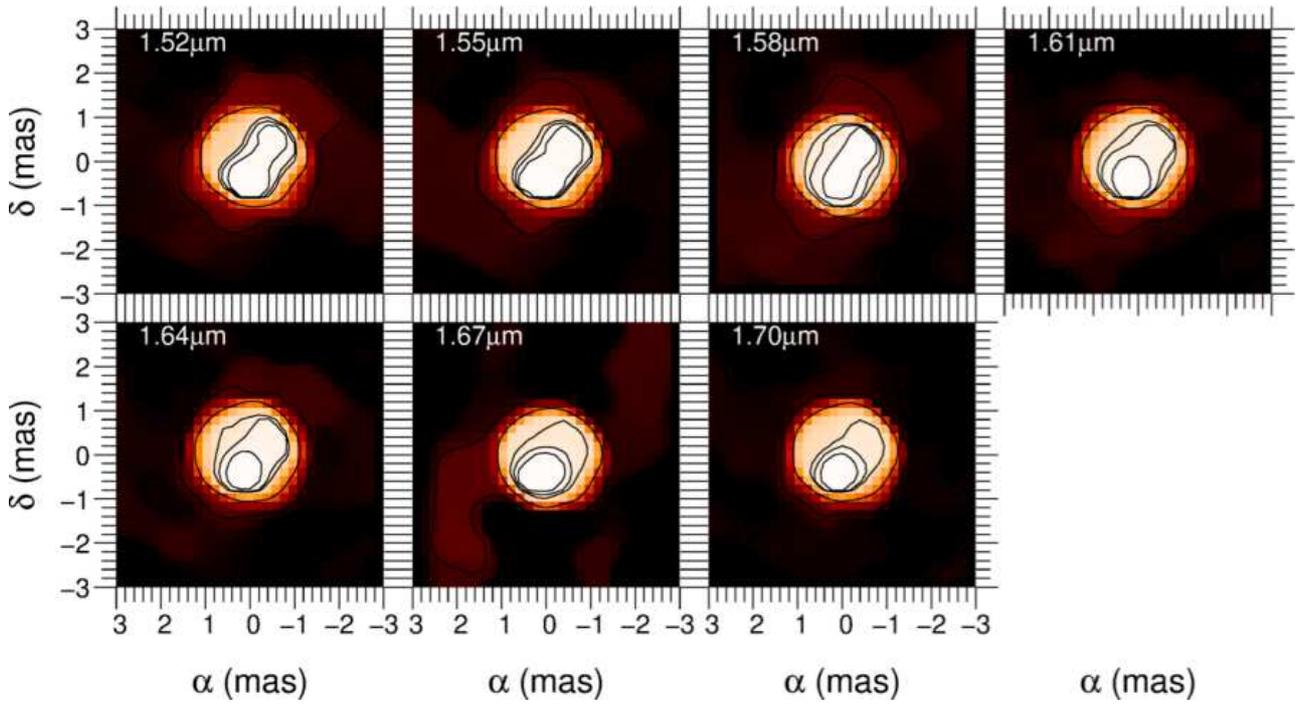} 
      \caption{Wavelength dependence of the reconstructed images. The intensity is normalised between [0, 1], and the contour lines are the same as in Fig.~\ref{im1}. These images have not been convolved with the interferometric beam.}
        \label{img_lambda}
\end{figure*}  

\section{How robust are our images?}
\begin{figure}
   \centering
    \includegraphics[height=1.0\hsize,angle=-90]{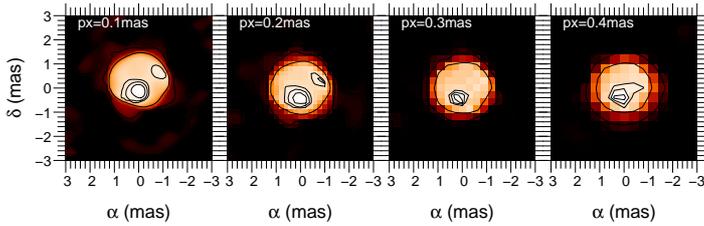} 
      \caption{Dependence of image appearance on the value of pixel size. The intensity is normalised between [0, 1] and the contour lines are the same as in Fig.~\ref{im1}. These images were not convolved with the interferometric beam.}
        \label{img_dep_px}
\end{figure}

\begin{figure}
   \centering
    \includegraphics[height=1.0\hsize,angle=-90]{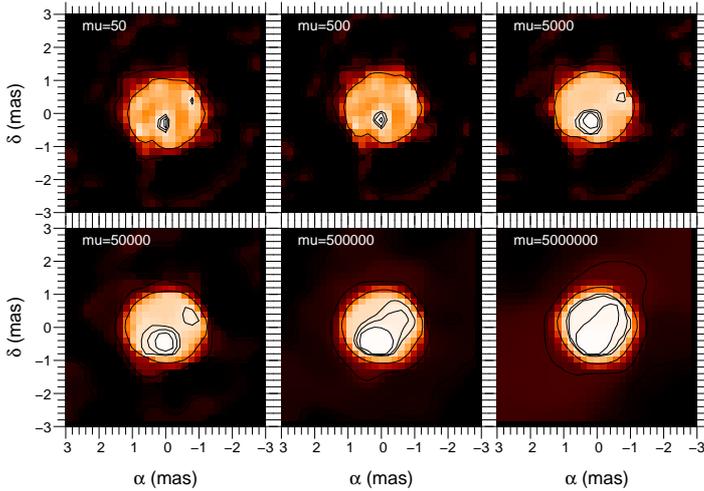} 
      \caption{Dependence of image appearance on the value of the hyperparameter $mu$. The intensity is normalised between [0, 1] and the contour lines are the same as in Fig.~\ref{im1}. These images were not convolved with the interferometric beam.}
        \label{img_dep_mu}
\end{figure}

\begin{figure}
   \centering
    \includegraphics[height=1.0\hsize,angle=-90]{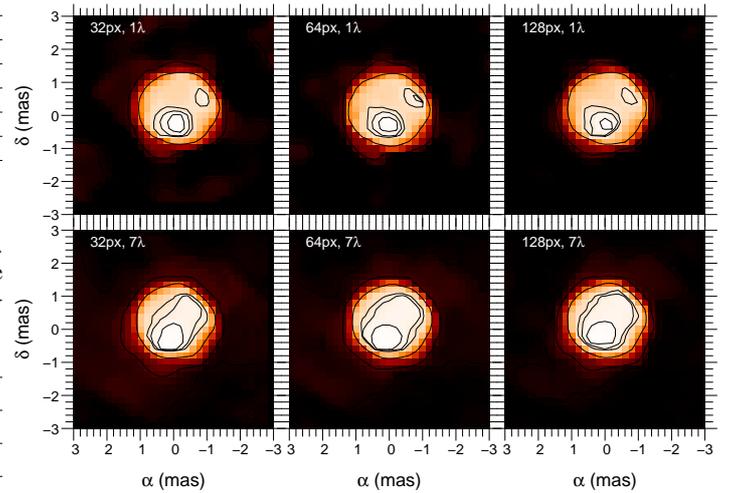} 
      \caption{Dependence of image appearance on the FOV. \emph{Top:} grey images (i.e. reconstructed with all $\lambda$ together). \emph{Bottom:} chromatic image computed with the median of all $\lambda$-reconstructed images. The intensity is normalised between [0, 1] and the contour lines are the same as in Fig.~\ref{im1}. These images were not convolved with the interferometric beam.}
        \label{img_dep_FOV}
\end{figure}

We present here a subset of all the images that were reconstructed on CL~Lac with the MIRA software to illustrate the robustness of the features discussed in this work. Image reconstruction with optical interferometry is known to have flaws due to the sparse sampling of the measurements, and also due to the intrinsically non-linear properties of the available interferometric observables (i.e. squared visibilities and closure phases). Therefore, we tested several of the input parameters of MIRA code, namely the pixel size (Fig.~\ref{img_dep_px}), the value of the hyperparameter $\mu$ (Fig.~\ref{img_dep_mu}), and the FOV (i.e. changing the number of pixels, Fig.~\ref{img_dep_FOV}). 

Figure~\ref{img_dep_px} shows that the image behaviour does not change drastically as a function of the pixel size: the disc of the star is the same size, and the bright area in the south-east is always located in the same place. We chose a pixel size of 0.2 mas. Then, Fig.~\ref{img_dep_mu} displays that too low $\mu$ values ($\leq10000$) return inhomogenous images, while too large values ($\geq100000$) tend to flatten the surface structures. We selected a median value of $\mu=50000$. Finally, Fig.~\ref{img_dep_mu} indicates that changing the image FOV does not affect its appearance.

\begin{figure}[!h]
   \centering
    \begin{tabular}{cc}
    \includegraphics[width=0.5\hsize]{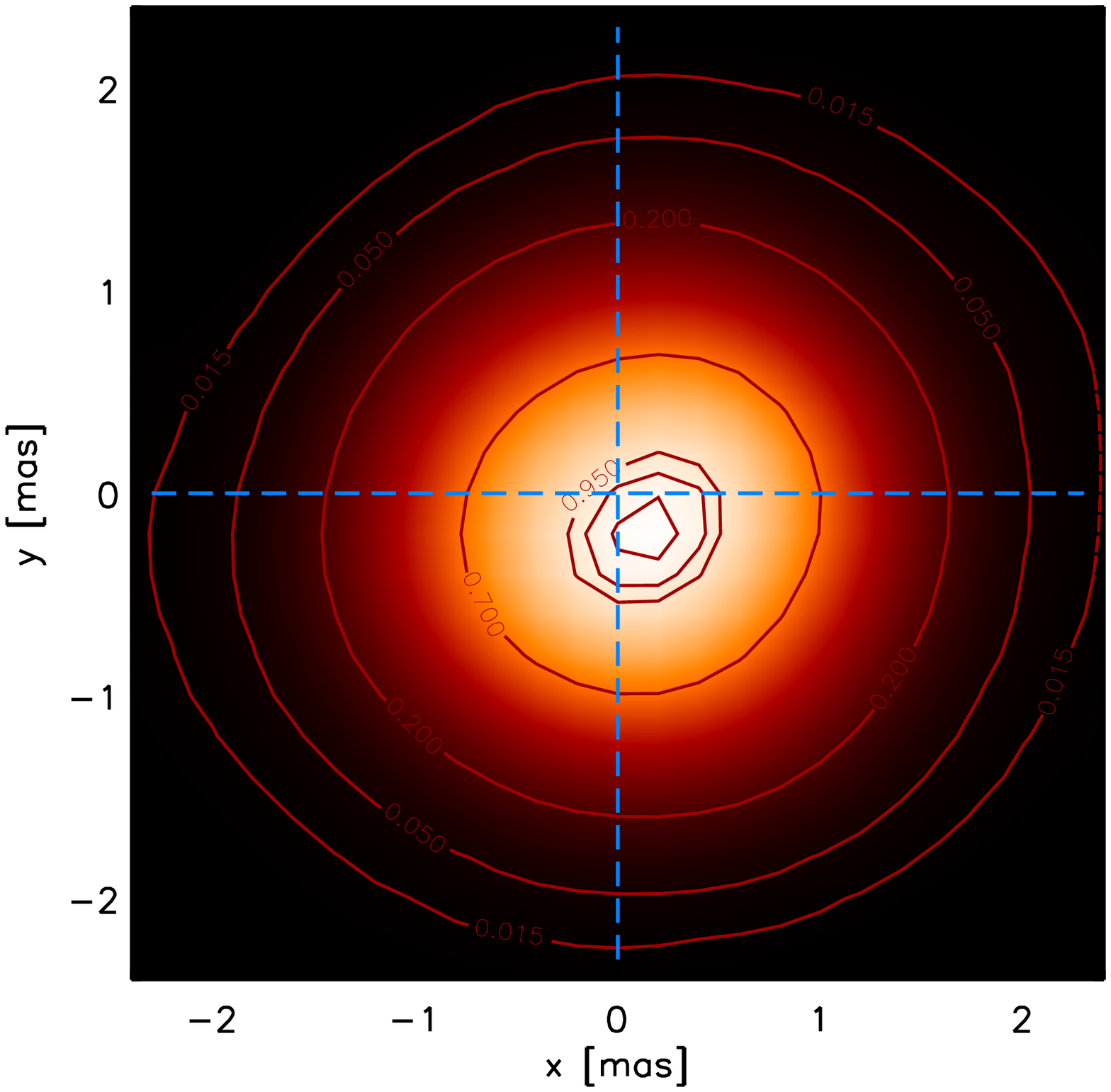} 
    \includegraphics[width=0.5\hsize]{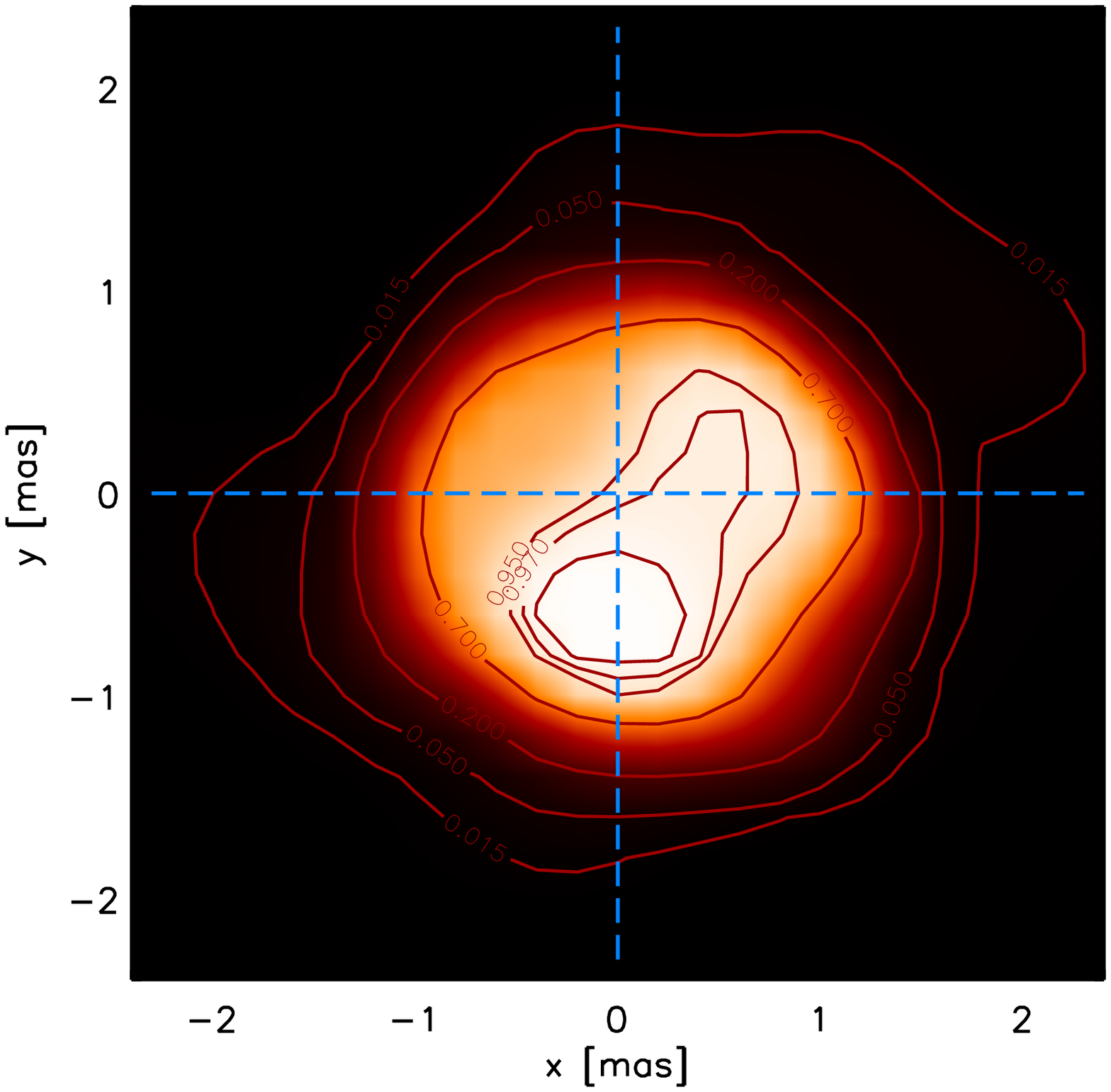} 
         \end{tabular}
      \caption{CL Lac reconstructed images convolved with an interferometric beam of FWHM $\lambda/B$ (left panel) and FWHM $\lambda/4B$ (right panel). The intensity is normalised between [0, 1] and the contour lines are the same as in Fig.~\ref{im1}.}
        \label{bis}
\end{figure}

Moreover, it has to be noted that the resolution of images reconstructed with MiRA is not necessarily $\lambda/B$, where $B$ is the largest baseline of the interferometric beam. Depending on the signal-to-noise and UV-plane coverage of the data, a degree of super-resolution (up to a factor of 2-3) is possible. In Fig.~\ref{bis}, we show two examples at 1.52 $\mu$m with other convolutions, namely an interferometric beam of FWHM $\lambda/(B)$ and $\lambda/(4B)$. These images have to be compared with those in Fig.~\ref{im1}.

 \begin{figure}
   \centering
    \includegraphics[height=1.0\hsize,angle=-90]{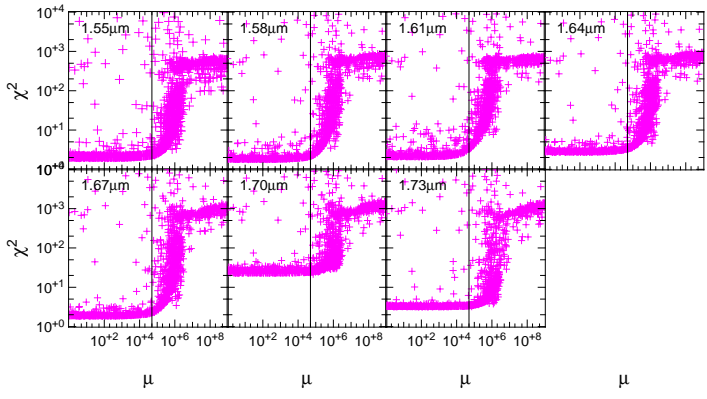} 
      \caption{L-curve plot: data term ($\chi^2$) as a function of the hyperparameter ($\mu$, all other image reconstruction parameters are random: pixel size, number of pixels, prior size, number of iterations).}
        \label{Lcurve_app}
\end{figure}

\end{appendix}

\end{document}